# Switchable Chern insulator, isospin competitions and charge density waves in rhombohedral graphene moiré superlattices

Jian Zheng[1†], Size Wu[1†], Kai Liu[1], Bosai Lyu[1], Shuhan Liu[1], Yating Sha[1], Zhengxian Li[1], Kenji Watanabe[2], Takashi Taniguchi[3], Jinfeng Jia[1,4], Zhiwen Shi[1], Guorui Chen[1*]

[1] Key Laboratory of Artificial Structures and Quantum Control (Ministry of Education), School of Physics and Astronomy, Shanghai Jiao Tong University, Shanghai 200240, China.

[2] Research Center for Electronic and Optical Materials, National Institute for Materials Science, 1-1 Namiki, Tsukuba 305-0044, Japan.

[3] Research Center for Materials Nanoarchitectonics, National Institute for Materials Science, 1-1 Namiki, Tsukuba 305-0044, Japan.

[4] Tsung-Dao Lee Institute, Shanghai Jiao Tong University, Shanghai, China.

†These authors contributed equally to this work.

*Correspondence to: chenguorui@sjtu.edu.cn

**Graphene-based moiré superlattices provide a versatile platform for exploring novel correlated and topological electronic states, driven by enhanced Coulomb interactions within flat bands. The intrinsic tunability of graphene's multiple degrees of freedom enables precise control over these complex quantum phases. In this study, we observe a range of competing phases and their transitions in rhombohedrally stacked hexalayer graphene on hexagonal boron nitride (*r*-6G/hBN) moiré superlattices. When electrons are polarized away from the moiré superlattice, we firstly identify a Chern insulator with reversible Chern numbers at *v* = 1 (one electron per moiré cell), attributed to the competition between bulk and edge magnetizations. Then, we detect transitions between three distinct insulating states at *v* = 2, driven by vertical displacement field *D* and vertical magnetic field *B*$_\perp$. These insulating phases are distinguished as spin-antiferromagnetic, spin-polarized, and valley-polarized insulators, based on their responses to parallel and perpendicular magnetic fields. When electrons are**

**polarized toward the moiré superlattice, in a device with large twist angle, insulating states appear at $v = 1/3$ and $2/3$ at zero magnetic field, and $v = 1/2$ in a magnetic field. Our findings reveal a rich interplay of charge, isospin, topology and magnetic field in rhombohedral graphene moiré superlattices.**

The exploration of correlated electronic states in solid-state systems reveals a diverse array of emergent phenomena, driven by complex interactions among electrons. Flat band systems are particularly notable in this context, as they significantly enhance Coulomb interactions by drastically reducing electron kinetic energies[1,2]. Moiré superlattices, created by stacking two-dimensional materials, provide a novel platform for engineering flat bands with multiple tunable parameters, including chemical composition, layer number, twisted angle, electrical and magnetic fields ($D$ and $B$)[2–9]. Rhombohedral graphene naturally hosts flat bands, as its energy dispersion follows $E \sim k^N$, where $N$ represents the number of layers[10–12]. As a result, Coulomb interactions strengthen when the number of layers increases, leading to the emergence of novel quantum phases, including superconductors[13,14], symmetry-broken metals[13,15–18], layer antiferromagnetic insulators[17–21] and high order Chern insulators[22–24]. Moreover, moiré superlattice between rhombohedral graphene and hexagonal boron nitride (hBN) can further narrow the bandwidth[25,26], leading to the emergence of new exotic phases that do not exist in crystalline graphene, such as correlated insulators at integer/fractional fillings[8,16,27–29] and integer/fractional Chern insulators[27–29].

The emergences of correlated and topological states in graphene are often accompanied by isospin (spin and valley) polarization. A notable example is the symmetry-broken metals observed in rhombohedral multilayer graphene, which displays stoner ferromagnetism with both spin and valley polarizations[13,16–18]. The magnetism of Chern insulators in graphene-based system is also believed to originate from valley polarization[30–32]. Moreover, the sign of the orbital Chern insulator and its magnetization can potentially be reversed when the Fermi level crosses the gap, as the total magnetization of an orbital Chern insulator arises from both the bulk and edge states[31,32].

Here, we fabricated two rhombohedrally stacked hexalayer graphene/hBN ($r$-6LG/hBN) moiré superlattice devices with different moiré periods (device I: $\theta = 0.51°$, device II: $\theta = 1.29°$). In device I, on the positive vertical displacement field (+$D$) side, we identified a Chern insulator at filling $v = 1$, where the sign of Chern number can be tuned by doping: negative ($C = -1$) under weak $p$-doping and positive ($C = +1$) under weak $n$-doping. Then, at filling $v = 2$, we observed transitions between insulating phases from an antiferromagnetic insulator to a spin polarized insulator, and finally to a valley polarized insulator, driven by $D$ and vertical magnetic field ($B_\perp$). The isospin orders in these insulating states are closely tied to symmetry breaking in crystalline $r$-6LG. In device II, on the negative vertical displacement field (-$D$) side, charge density waves (CDW) insulating phases with different spin textures were observed at fractional fillings of 1/3 and 2/3. Under a vertical magnetic field, a stripe phase appears at $v = 1/2$.

Figure. 1a depicts a schematic cross-sectional view of the device, showcasing a dual gate structure that enables independent tuning of carrier density $n$ and the vertical displacement field $D$. To eliminate any possible moiré superlattice between $r$-6LG and top hBN, monolayer tungsten disulfide ($WS_2$) is used as a spacer. Consequently, the moiré superlattice arises solely from the alignment between $r$-6LG and bottom hBN. The optical image of device I is shown in Extended Data Fig. 1, with a moiré period of $L = 13.4$ nm and a twist angle of $\theta = 0.51°$, as confirmed by the carrier concentration of the resistance peaks at integer fillings, and the carrier concentration is confirmed by the landau fan diagram (Extended Data Fig. 2). Figure 1b shows the four-probe resistance $R_{xx}$ of device I as functions of $v$ and $D$. Line plots of $R_{xx}$ at different $D$ is shown in Fig. 1c. As $D$ is applied in either direction, the resistance at $v = 0$ increases due to band gap opening. Upon electron doping, four resistance peaks are well developed on the -$D$ side at integer fillings of $v = 1, 2, 3$, and 4, whereas on the +$D$ side, resistance peaks appear only at $v = 1$ and 2. The resistance peaks at $v = 4$ corresponds to full filling of the first conduction miniband, while other insulating peaks at integer fillings from 1 to 3 arises from correlations. The asymmetry between +$D$ and -$D$ side stems from different strength of moiré potential experienced by the electrons in the top and bottom graphene layers. On the -$D$ side, the electrons are closer to the moiré superlattice and experiences stronger moiré modulation, while on the +D side, electrons are polarized toward the top graphene layer, moving away from the moiré superlattice. This is further evidenced by clear phase boundaries between isospin-polarized metals on the +D side. While the phase boundaries are almost absent due to strong moiré modulations on the -D side. Notably, no insulating peaks are observed on the hole-doping side within the range of D explored in our experiment. The electron-hole asymmetry can be explained by the wider bandwidth on the hole side relative to the electron side in the r-6LG/hBN moiré superlattice[33–35].

On the +D side, the $v = 1$ correlated insulator is predicted to be topological Chern insulator with a Chern number of $C = 1$[33–35]. Interestingly, however, we initially observe a $C = -1$ Chern insulator at this filling in our device. Figure. 2a and b present the longitudinal resistance $R_{xx}$ and Hall resistance $R_{xy}$ as functions of $v$ and $D$ at $B_\perp = 1$T. On either doping side of the $v = 1$ insulating peak, dips in $R_{xx}$ are evident, as indicated by the dashed lines in Fig. 2a. In the same region (indicated by dashed lines in Fig 2b), large Hall signals are observed, but with different signs for hole- and electron-doping (Fig. 2b and Extended Data Fig. 4). Fixing the doping on the hole side of the $v = 1$ insulator at $D = 0.65$ V/nm, we observe a pronounced anomalous Hall signal in $R_{xy}$ upon sweeping $B_\perp$, as shown in Fig. 2e. The anomalous Hall resistance reaches 15 kΩ at zero magnetic field, and from the sign of $R_{xy}$ at positive and negative magnetic fields, we determine the Chern number to be $C = -1$. When the moiré superlattice is slightly doped on the electron side of the $v = 1$ insulator, an anomalous Hall signal with same sign is detected near zero magnetic field (Fig. 2f). With increasing magnetic field, $R_{xy}$ reverses its sign, and $R_{xy}$ reach the quantized value of h/e² at 2.5T, corresponding to the predicted $C = 1$ Chern insulator. The development of the $C = ±1$ Chern insulator is more clearly illustrated in the Landau fan diagram (Figs. 2c and 2d), where both Chern insulating states ($C = ±1$) disperse with increasing $B_\perp$, consistent with the Streda's formula (marked by the two black dashed lines). These observations suggest that the Chern insulator here is $n$-dependent, and its sign can be reversed by tuning the doping. Same measurements at $D = 0.62$ V/nm and $D = 0.7$ V/nm reveal the same sign-reversal behavior (Extended Data Fig. 5). As shown in Figs. 2a and 2b, Chern insulating states with opposite signs start to appear at $D = 0.55$ V/nm, and persists up to 0.73 V/nm, the highest $D$ applied in our experiment.

The doping-induced sign reversal of Chern number $C$ is related to the sign change in total magnetization $M$. In an orbital Chern insulator, the total magnetization $M$ arises from both bulk $M_{bulk}$ and edge state $M_{edge}$. If $M_{edge}$ is large enough, then $M$ can change sign as the Fermi level crosses the gap[23,31,32]. The evolution of Chern insulator in magnetic field can be understood by the relationship between $C$ and $M$. If $M$ does not change sign across the gap and $C_{\pm}M_{\pm} > 0$ ($C_{\pm}M_{\pm} < 0$), the magnetic field stabilizes the positive (negative) $C$ for positive $B$ and the negative (positive) $C$ for negative B. In the $r$-6LG/hBN superlattice, $M$ changes sign across the gap, making both signs ($\pm$) of the Chern insulator robust in a magnetic field. Detailed illustrations for the three cases have been discussed in the supplementary material of Ref 31 and are also provided in Extended Data Fig. 6.

On the $+D$ side, the correlated insulator at $v = 2$ is generally believed to be topological trivial because it does not occupy a pure valley. Indeed, we do not see any anomalous hall signal near $v = 2$. However, as shown in Figs. 1b and c, there are two insulating regions at $v = 2$: one from 0.1 V/nm to 0.18 V/nm, the other from 0.4 V/nm to 0.74 V/nm. Figure. 3a shows a line cut at $v = 2$ along $D$ axis, where the insulating states are labeled as I and II, respectively. This $D$-induced insulator-metal-insulator transition is attributed to isospin competitions. In order to determine the isospin flavors in these two insulating states, we applied perpendicular and parallel magnetic fields $B_\perp$ and $B_\parallel$ with respect to graphene plane, because valley only responds to $B_\perp$ and spin equally responds to $B_\perp$ and $B_\parallel$. The resulting $D$ - $B_\perp$ and $D$ - $B_\parallel$ maps of $R_{xx}$ are shown in Fig. 3b and c, respectively. Both insulators I and II are suppressed with increasing $B_\perp$, with insulator I persisting up to $B_\perp = 2.1$ T and insulator II up to $B_\perp = 6.5$ T. However, the two insulators exhibit opposite responses to $B_\parallel$: insulator I is suppressed, while insulator II is enhanced. Remarkably, a new high-resistance state labeled as III emerges at higher $D$ as $B_\perp$ increases, which is separated by a phase boundary from the insulator II.

To quantitatively compare these insulating states, we varied the temperature and estimated their transport gap. Fig. 3d shows the evolution of the transport gaps of the three insulating states as a function of $B_\parallel$ or $B_\perp$. At zero magnetic field, the gap size of insulator I is approximately 2.9 meV ($D = 0.14$ V/nm), and that of insulator II is about 0.88 meV ($D = 0.48$ V/nm). As $B_\parallel$ increases, the gap of the insulator I decreases linearly, following the Zeeman energy $g_I \mu_B B_\parallel$, with an estimated g-factor of $g_I \approx -2$. In contrast, the gap of the insulator II increases linearly with $B_\parallel$, also following the Zeeman energy $g_{II} \mu_B B_\parallel$, with an estimated g-factor of $g_{II} \approx 2$. These observations suggest that insulator I is an antiferromagnetic insulator (AFI), and the insulator II is a spin polarized insulator (SPI). For the insulator III, the gap size (at $D = 0.60$ V/nm) increases linearly with $B_\perp$, following $g_{III} \mu_B B_\perp$ with an estimated g-factor of $g_{III} = 9$. This large g factor suggests that valley plays a more substantial role than spin in insulator III. Thus, insulator III is identified as a valley polarized insulator (VPI).

These insulating phases at $v = 2$ is schematically illustrated in Fig. 3d, which shows the isospin ground states for the three insulators. We further compared these insulators at $v = 2$ with the isospin symmetry-broken metals in crystalline *r*-6LG (Extended Data Fig. 3) without moiré. We find that the SPI state is situated in the spin polarized half-metal region, the VPI state is in the valley polarized quarter-metal region, and the AFI state lies at the phase boundary between normal metal and half-metal regions. This indicates that the isospin polarization of the correlated insulators in the moiré superlattice is closely related to the spontaneously symmetry-broken states in crystalline *r*-6LG.

We measured a second device with smaller moiré period $L = 8.93$ nm and larger twist angle $\theta = 1.29°$. The transport data of this device, shown in Fig. 4a, reveal several differences from the first device. First, at $v = 0$, the layer antiferromagnetic (LAF) insulator at $D = 0$ V/nm survives, similar to that observed in crystalline *r*-6LG without moiré (Extended Data Fig. 3). Second, the AFI state (insulator I) at $v = 2$ is absent. Third, other integer filling insulators in this device emerge at higher $D$ values, and the phase boundaries between isospin-polarized metals are distinct for both directions of $D$.

These differences are due to the weaker moiré modulations in the device with larger twist angle.

Besides these notable differences, there is a tilted stripe-shaped high resistance region at -$D$ side (-0.95V/nm ~ -0.75V/nm) between $v = 0$ and 1. Figure. 4b presents a zoomed-in view of this region, where two resistance peaks at fractional fillings $v = 1/3$ and 2/3 are clearly resolved. These two insulating states are referred to charge density wave (CDW). The transport gaps of these two CDW states are about 0.85 meV and 0.57 meV, respectively, based on the temperature-dependent measurements (Fig. 4c). Interestingly, under the influence of increasing $B_\perp$, these two CDW states shifts to smaller |$D$|. Starting from $B_\perp = 5$ T, a new resistance peak at $v = 1/2$ starts to appear and is further prominent under higher $B_\perp$ (Fig. 4d). The state at $v = 1/2$ breaks the $C_3$ symmetry of the moiré superlattice, and is referred as a stripe phase insulator. We extracted the $D$ and $B_\perp$ values of the maximum resistances of these three insulating states at fractional fillings and plotted them in Fig. 4e. They all exhibit roughly linear relations, and we note that the slope of 1/3 and 2/3 states are almost the same to be 0.02 V/(nm · T). While the slope of 1/2 state is about 0.014 (V/nm · T). The effect of $B_\perp$ is possibly related to the interplay between magnetic flux and the CDW pattern. The magnetic flux is described as $\Phi = B_\perp*A$, where A is the area of CDW unit cell. For $v = 1/3$ (2/3) and 1/2, A is two and three times of moiré superlattice unit cell, respectively. Therefore, the ratio of the magnetic flux between 1/3(2/3) and 1/2 is 3 : 2 (=1.5), which roughly agrees with the ratio of the slope (0.02 : 0.014 = 1.43) in Fig. 4e. Under increasing $B_\parallel$, the 1/3 and 2/3 CDW states exhibit opposite behaviors. At $D = -0.85$ V/nm, the 1/3 CDW state is well-developed while 2/3 state is absent (Fig. 4b). As $B_\parallel$ increases, the resistance of the 1/3 state decreases, eventually disappearing around $B_\parallel = 10$ T (Fig. 4g), while the 2/3 state appears near 5 T. At $D = -0.88$ V/nm, the 2/3 state is well-developed at zero magnetic field (Fig. 4b) and its resistance increases with $B_\parallel$, whereas the 1/3 state remain absent at $B_\parallel = 12$ T (Fig. 4f). This contrasting dependence on $B_\parallel$ suggests the 1/3 and 2/3 CDW states distinct spin textures.

At last, we note that the LAF state is absent in device I and present in device II. We attribute the absence of LAF state in device I to the moiré-induced imbalance between the top and bottom graphene layers. Previous investigations have shown that graphene/hBN moiré superlattice ($L = 13.5$ nm) can create large band gap about 25 meV at $v = 0$[36]. This energy scale agrees with the gap size of LAF state in crystalline $r$-6LG, which is about 27 meV (Extended Data Fig. 3d).

In summary, we report the observation of a variety of topological and correlated phase in $r$-6LG/hBN moiré superlattice on both sides of $D$. On the $+D$ side, we identify a switchable Chern insulator at filling $v = 1$. At $v = 2$, we observe three insulating phases with isospin competitions, which can be controlled by $D$ and $B$. Additionally, correlated insulators at fractional fillings of 1/3, 1/2 and 2/3 emerge at $-D$ side. These findings unveil a wealth of correlated and topological phases with multiple competing orders and tuning parameters in rhombohedral multilayer graphene moiré superlattices. Our results showcase moiré superlattice as a condensed matter quantum simulator, offering a versatile platform for the realization and precise tuning of diverse quantum states.

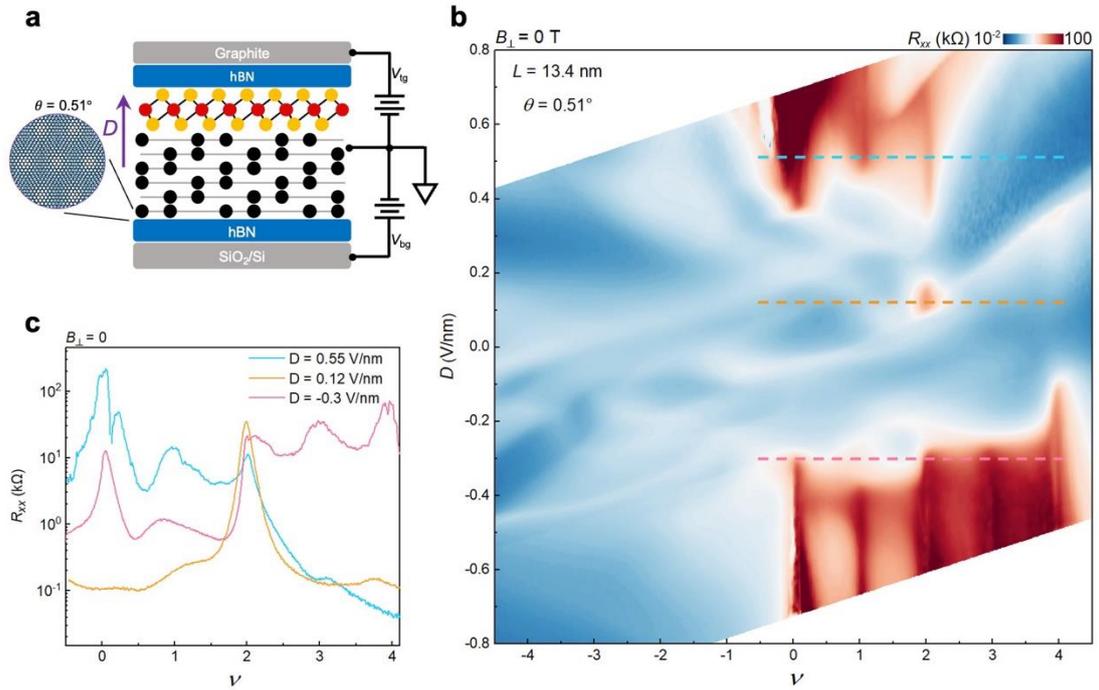

**Fig. 1 | Device structure and transport characterization of *r*-6LG/hBN moiré superlattice with $\theta = 0.51°$. a,** Schematic cross-sectional view of the device. The *r*-6LG is encapsulated in hBN flakes, forming a moiré superlattice with a period of $L = 13.4$ nm and $\theta = 0.51°$ between *r*-6LG and bottom hBN. Monolayer $WS_2$ is used as a spacer to exclude possible effects from the moiré superlattice between *r*-6LG and top hBN. **b,** Color plot of the four-probe resistance $R_{xx}$ as functions of $v$ and $D$ at zero magnetic field. **c,** $R_{xx}$ as a function of $v$ at $D = 0.55$ V/nm, 0.12 V/nm, and –0.3 V/nm, corresponding to the dashed lines in Fig. 1b.

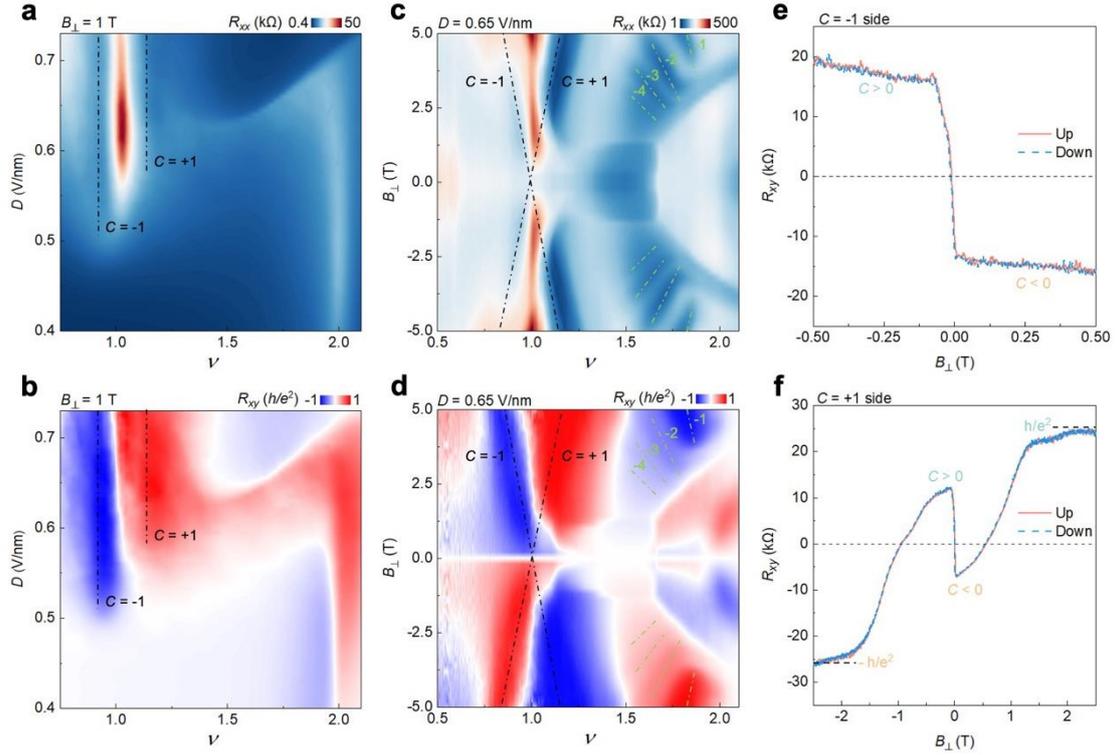

**Fig. 2 | Electrically switchable Chern insulators. a, b,** Color plots of $R_{xx}$ **(a)** and $R_{xy}$ **(b)** as functions of $v$ and $D$ at $B_\perp = 1$ T. Dashed lines indicates the Chern insulator at hole ($C = -1$) and electron ($C = +1$) doping sides. **c, d,** Landau fan diagrams of symmetrized $R_{xx}$ **(c)** and anti-symmetrized $R_{xy}$ **(d)** as functions of $v$ and $B_\perp$ at $D = 0.65$ V/nm. The Chern insulators with $C = \pm 1$ at $v = 1$ are highlighted by black dashed lines, which follows the Streda formula. Landau level-induced quantum Hall states are labeled by green dashed lines with filling factors of -1, -2, -3, and -4. **e, f,** Anomalous Hall signals $R_{xy}$ at $C = -1$ ($v = 0.95$) side **(e)** and $C = +1$ ($v = 1.05$) side **(f)**.

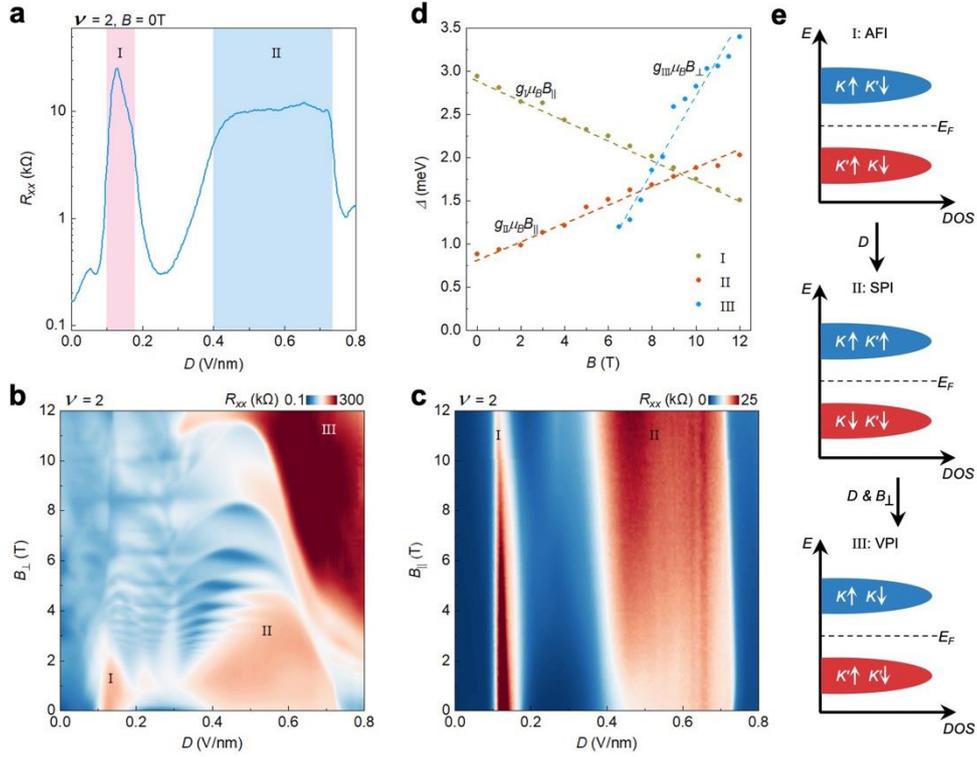

**Fig. 3 | Isospin competitions in the correlated insulators at $v = 2$. a,** $R_{xx}$ as a function of $D$ at $v = 2$. The two insulating states at different $D$ are labelled as I and II, respectively. **b, c,** Color plots of $R_{xx}$ at $v = 2$ as functions of $D$ and $B_\perp$ (**b**) and $B_\parallel$ (**c**). $R_{xx}$ of insulator I decreases with both $B_\perp$ and $B_\parallel$. In contrast, $R_{xx}$ of insulator II decreases with $B_\perp$ but increases with $B_\parallel$. As $B_\perp$ increases, a new insulating phase gradually develops at higher $D$, which is labeled as III. **d,** Transport gaps $\Delta$ of the three insulators as a function of $B_\parallel$ (for I and II) or $B_\perp$ (for III). **e,** Schematic illustration of the isospin flavor evolution for the three insulating states. The *x*-axis represents the density of states (*DOS*), the *y*-axis represents energy, $E_F$ denotes the Fermi level, $K$ and $K'$ represents two valleys, and ↑ and ↓ represents two spins.

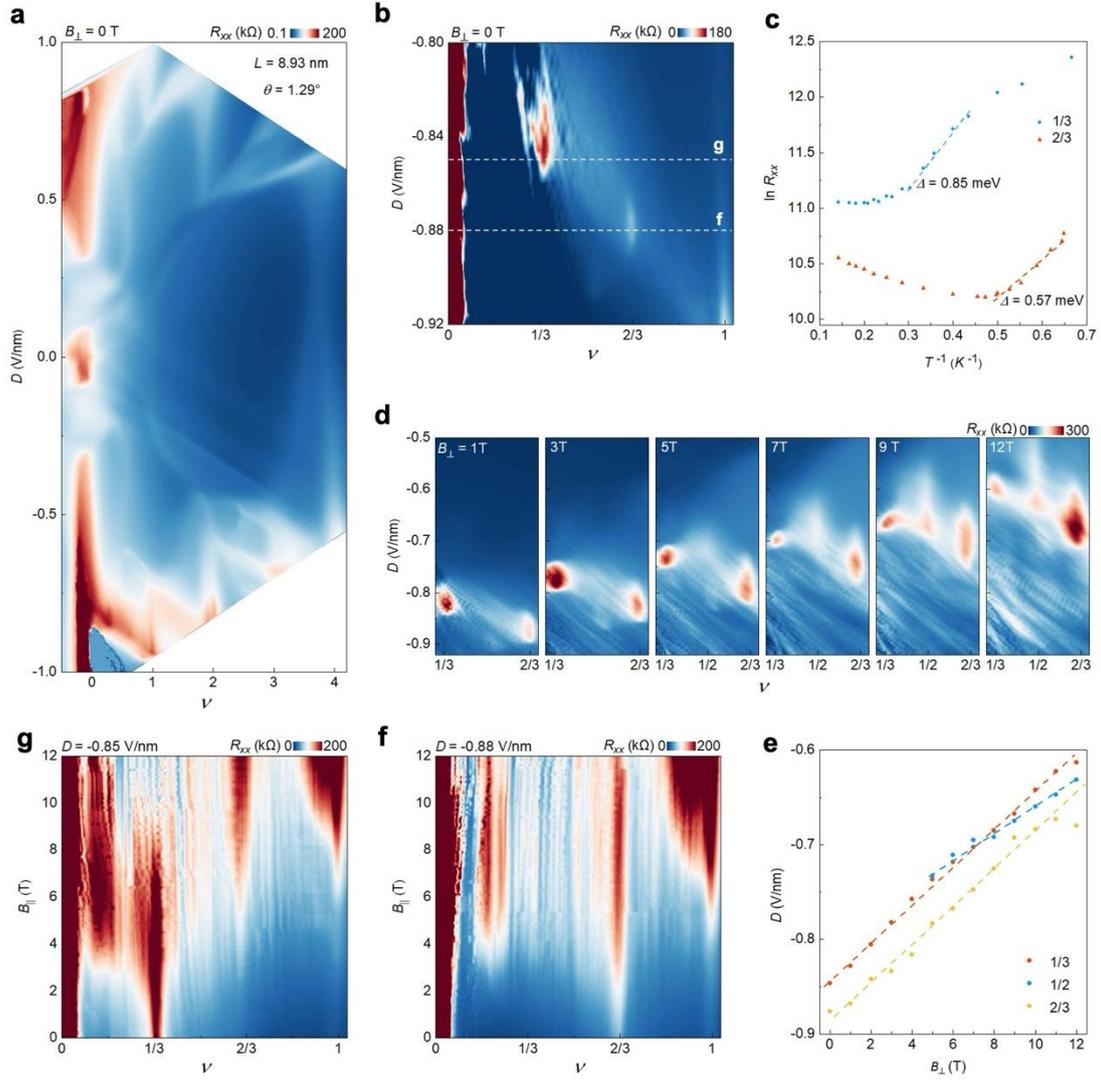

**Fig. 4 | Insulators at fractional fillings in the *r*-6LG/hBN moiré superlattice with $\theta = 1.29°$. a,** Color plot of $R_{xx}$ as functions of $v$ and $D$. This device has a moiré period of $L = 8.93$ nm, $\theta = 1.29°$. **b,** Zoomed in color plot of $R_{xx}$ at $-D$ and $0 < v < 1$. CDW states develop at fractional fillings 1/3 and 2/3. **c,** Arrhenius plot of $\ln R_{xx}$ versus $T^{-1}$ at 1/3 and 2/3 CDW states. The transport gap is extracted from the thermal activation equation $R_{xx} \propto \exp(-\Delta/2k_BT)$, with the gaps estimated to be approximately 0.85 meV and 0.57 meV for the 1/3 and 2/3 state, respectively. Dashed lines represent the linear regions used for estimating the gaps. **d,** Evolutions of CDW states under increasing $B_\perp$. As $B_\perp$ increases, the CDW states shift to smaller $D$. At about 5T, a stripe phase appears at $v = 1/2$. **e,** $D$ as a function of $B_\perp$ for three insulating states. **g and f,** Color plots of $R_{xx}$ as functions of $v$ and $B_\parallel$ at $D = -0.85$ and $-0.88$ V/nm, respectively.

# References

1. Paschen, S. & Si, Q. Quantum phases driven by strong correlations. *Nat. Rev. Phys.* **3**, 9–26 (2020).
2. Nuckolls, K. P. & Yazdani, A. A microscopic perspective on moiré materials. *Nat. Rev. Mater.* **9**, 460–480 (2024).
3. Li, T. *et al.* Quantum anomalous Hall effect from intertwined moiré bands. *Nature* **600**, 641–646 (2021).
4. Bistritzer, R. & MacDonald, A. H. Moiré bands in twisted double-layer graphene. *Proc. Natl. Acad. Sci.* **108**, 12233–12237 (2011).
5. Cao, Y. *et al.* Correlated insulator behaviour at half-filling in magic-angle graphene superlattices. *Nature* **556**, 80–84 (2018).
6. Cao, Y. *et al.* Unconventional superconductivity in magic-angle graphene superlattices. *Nature* **556**, 43–50 (2018).
7. Cao, Y. *et al.* Tunable correlated states and spin-polarized phases in twisted bilayer–bilayer graphene. *Nature* **583**, 215–220 (2020).
8. Chen, G. *et al.* Evidence of a gate-tunable Mott insulator in a trilayer graphene moiré superlattice. *Nat. Phys.* **15**, 237–241 (2019).
9. Regan, E. C. *et al.* Mott and generalized Wigner crystal states in WSe2/WS2 moiré superlattices. *Nature* **579**, 359–363 (2020).
10. Yelgel, C. Electronic Structure of ABC-stacked Multilayer Graphene and Trigonal Warping:A First Principles Calculation. *J. Phys. Conf. Ser.* **707**, 012022 (2016).
11. Koshino, M. & McCann, E. Trigonal warping and Berry's phase N $\pi$ in ABC-stacked multilayer graphene. *Phys. Rev. B* **80**, 165409 (2009).
12. Min, H. & MacDonald, A. H. Electronic Structure of Multilayer Graphene. *Prog. Theor. Phys. Suppl.* **176**, 227–252 (2008).
13. Zhou, H. *et al.* Isospin magnetism and spin-polarized superconductivity in Bernal bilayer graphene. *Science* **375**, 774–778 (2022).
14. Zhou, H., Xie, T., Taniguchi, T., Watanabe, K. & Young, A. F. Superconductivity in rhombohedral trilayer graphene. *Nature* **598**, 434–438 (2021).
15. Han, T. *et al.* Orbital multiferroicity in pentalayer rhombohedral graphene. *Nature* **623**, 41–47 (2023).
16. Zhou, H. *et al.* Half- and quarter-metals in rhombohedral trilayer graphene. *Nature* **598**, 429–433 (2021).
17. Han, T. *et al.* Correlated insulator and Chern insulators in pentalayer rhombohedral-stacked graphene. *Nat. Nanotechnol.* 19. 181-187 (2024).
18. Liu, K. *et al.* Spontaneous broken-symmetry insulator and metals in tetralayer rhombohedral graphene. *Nat. Nanotechnol.* **19**, 188–195 (2024).
19. Weitz, R. T., Allen, M. T., Feldman, B. E., Martin, J. & Yacoby, A. Broken-Symmetry States in Doubly Gated Suspended Bilayer Graphene. *Science* **330**, 812–816 (2010).
20. Feldman, B. E., Martin, J. & Yacoby, A. Broken-symmetry states and divergent resistance in suspended bilayer graphene. *Nat. Phys.* **5**, 889–893 (2009).

21. Velasco, J. *et al.* Transport spectroscopy of symmetry-broken insulating states in bilayer graphene. *Nat. Nanotechnol.* **7**, 156–160 (2012).
22. Geisenhof, F. R. *et al.* Quantum anomalous Hall octet driven by orbital magnetism in bilayer graphene. *Nature* **598**, 53–58 (2021).
23. Sha, Y. *et al.* Observation of a Chern insulator in crystalline ABCA-tetralayer graphene with spin-orbit coupling. *Science* **384**, 414–419 (2024).
24. Han, T. *et al.* Large quantum anomalous Hall effect in spin-orbit proximitized rhombohedral graphene. *Science* **384**, 647–651 (2024).
25. Park, Y., Kim, Y., Chittari, B. L. & Jung, J. Topological flat bands in rhombohedral tetralayer and multilayer graphene on hexagonal boron nitride moiré superlattices. *Phys. Rev. B* **108**, 155406 (2023).
26. Chittari, B. L., Chen, G., Zhang, Y., Wang, F. & Jung, J. Gate-Tunable Topological Flat Bands in Trilayer Graphene Boron-Nitride Moiré Superlattices. *Phys. Rev. Lett.* **122**, 016401 (2019).
27. Xie, J. *et al.* Even- and Odd-denominator Fractional Quantum Anomalous Hall Effect in Graphene Moiré Superlattices.
28. Chen, G. *et al.* Magnetic Field-Stabilized Wigner Crystal States in a Graphene Moiré Superlattice. *Nano Lett.* **23**, 7023–7028 (2023).
29. Lu, Z. *et al.* Fractional quantum anomalous Hall effect in multilayer graphene. *Nature* **626**, 759–764 (2024).
30. Chen, G. *et al.* Tunable correlated Chern insulator and ferromagnetism in a moiré superlattice. *Nature* **579**, 56–61 (2020).
31. Zhu, J., Su, J.-J. & MacDonald, A. H. Voltage-Controlled Magnetic Reversal in Orbital Chern Insulators. *Phys. Rev. Lett.* **125**, 227702 (2020).
32. Polshyn, H. *et al.* Electrical switching of magnetic order in an orbital Chern insulator. *Nature* **588**, 66–70 (2020).
33. Zhou, B., Yang, H. & Zhang, Y.-H. Fractional quantum anomalous Hall effects in rhombohedral multilayer graphene in the moireless limit and in Coulomb imprinted superlattice. Preprint at http://arxiv.org/abs/2311.04217 (2023).
34. Dong, J. *et al.* Anomalous Hall Crystals in Rhombohedral Multilayer Graphene I: Interaction-Driven Chern Bands and Fractional Quantum Hall States at Zero Magnetic Field. Preprint at http://arxiv.org/abs/2311.05568 (2023).
35. Dong, Z., Patri, A. S. & Senthil, T. Theory of fractional quantum anomalous Hall phases in pentalayer rhombohedral graphene moiré structures. Preprint at http://arxiv.org/abs/2311.03445 (2023).
36. Hunt, B. *et al.* Massive Dirac Fermions and Hofstadter Butterfly in a van der Waals Heterostructure. *Science* **340**, 1427–1430 (2013).

**Materials and Methods**

S1. Device fabrications

Graphene, WS$_2$, graphite, and hBN were mechanically exfoliated onto SiO$_2$ (285 nm)/Si substrates. The layer numbers of graphene and WS$_2$ were determined using optical contrast techniques. The stacking order of hexalayer graphene ($r$-6LG) was verified with a scanning near-field optical microscope (SNOM). To construct the hBN/WS$_2$/$r$-6LG/hBN moiré heterostructures, a dry transfer technique was employed, followed by a second SNOM characterization to confirm the stacking order of $r$-6LG. Subsequently, standard e-beam lithography, reactive ion etching, and metal evaporation were performed to fabricate the devices into a Hall bar geometry with one-dimensional edge contacts, as shown in Fig. S2.

S2. Transport measurement

Transport measurements were conducted in a 1.5K Oxford variable temperature insert (VTI) system. The sample resistances were measured using Stanford Research Systems SR860, SR830, and Guangzhou Sine Scientific Instrument OE1201 lock-in amplifiers with a small AC bias current of 5-10 nA at a frequency of approximately 17.777Hz, in combination with a 100 MΩ resistor. Gate voltages were applied using Keithley 2400 Source Meters. In our dual gate device, the top and bottom gate voltages ($V_t$, $V_b$) can independently tune the carrier density $n$ and the vertical displacement field $D$ in $r$-6LG/hBN Moiré superlattice: $n = (D_b + D_t)/e$, $D = (D_b - D_t)/2$, where $D_b = \varepsilon_b(V_b - V_b^0)/d_b$ and $D_t = \varepsilon_t(V_t - V_t^0)/d_t$ Here, $e$, $\varepsilon$ and $d$ represent the electron charge, dielectric constant and thickness of the dielectric layers respectively, $V_b^0$ and $V_t^0$ are effective offset voltages caused by intrinsic carrier doping.

S4. Symmetrization and anti-symmetrization of the magneto-transport data

We note that the longitudinal $R_{xx}$ and transverse $R_{xy}$ signals can mix during our measurements. The longitudinal signal $R_{xx}$ is symmetric while the transverse $R_{xy}$ signals is anti-symmetric with respect to magnetic field $B$. we can separate the two signals using standard symmetrization and anti-symmetrization procedures. The symmetrization and anti-symmetrization procedure is as follows:

$$R_{xx}(B) = \frac{R_{12,34}(B) + R_{12,34}(-B)}{2}$$
$$R_{xy}(B) = \frac{R_{12,45}(B) - R_{12,45}(-B)}{2}$$

where 1, 2, 3, 4, and 5 denote the contact numbers, with contacts 1-4 on one side of the sample channel and contact 5 on the opposite side. For example, $R_{12,34}(B)$ corresponds to a resistance measured by applying current from contact 1 to 2 and measuring the voltage between contact 3 and 4. Most of the figures presented in this paper are raw data without any processing, with the exception of those in **Fig. 2c-d and Extended Data Fig. 5**, which have been symmetrically and anti-symmetrically processed.

**Acknowledgments**

G.C. acknowledges support from National Key Research Program of China (grant nos. 2021YFA1400100 and 2020YFA0309000), NSF of China (grant nos. 12350005 and 12174248), and Yangyang Development Fund. K.W. and T.T. acknowledge support from the JSPS KAKENHI (Grant Numbers 20H00354, 21H05233 and 23H02052) and World Premier International Research Center Initiative (WPI), MEXT, Japan.

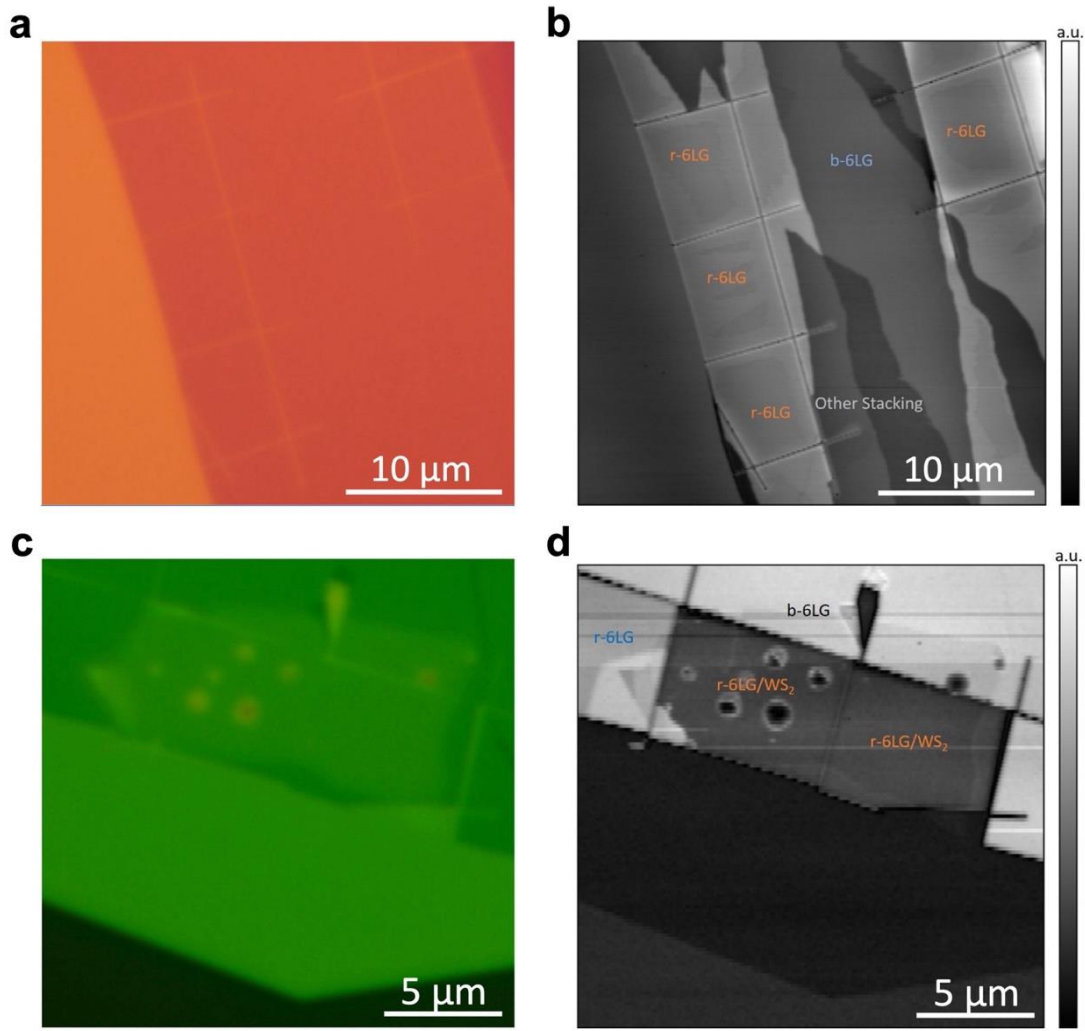

**Extended Data Fig. 1. Identification of *r*-6LG/hBN superlattice. a,** Optical image of the 6LG, with the *r*-6LG isolated by anodic-oxidation-assisted AFM cutting. **b,** SNOM image of the 6LG, showing three contrasts that indicate different stacking orders, *r*-6LG, Bernal-stacked 6LG (*b*-6LG) and other stacking. **c,** Optical image of the hBN/WS$_2$/*r*-6LG/hBN heterostructure prepared using the dry transfer method. **d,** SNOM image of the the hBN/WS$_2$/*r*-6LG/hBN heterostructure.

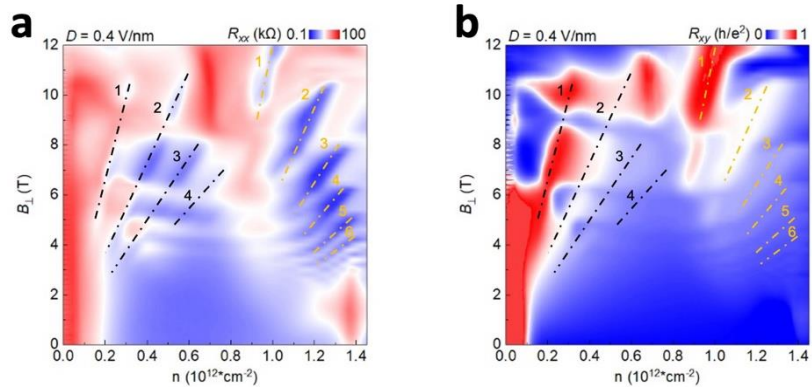

**Extended Data Fig. 2. Landau fan diagram at $D = 0.4$ V/nm in device 1. a, b,** Landau fan diagrams showing the longitudinal resistance $R_{xx}$ **(a)** and transverse resistance $R_{xy}$ **(b)** as functions of the carrier density $n$ and the magnetic field $B_\perp$ at $D = 0.4$ V/nm. Dashed lines and numbers (1, 2, 3, 4, …) indicate the landau levels. From the Landau fan diagram, the moiré period is estimated to be roughly $L = 13.43$ nm and $\theta = 0.51°$.

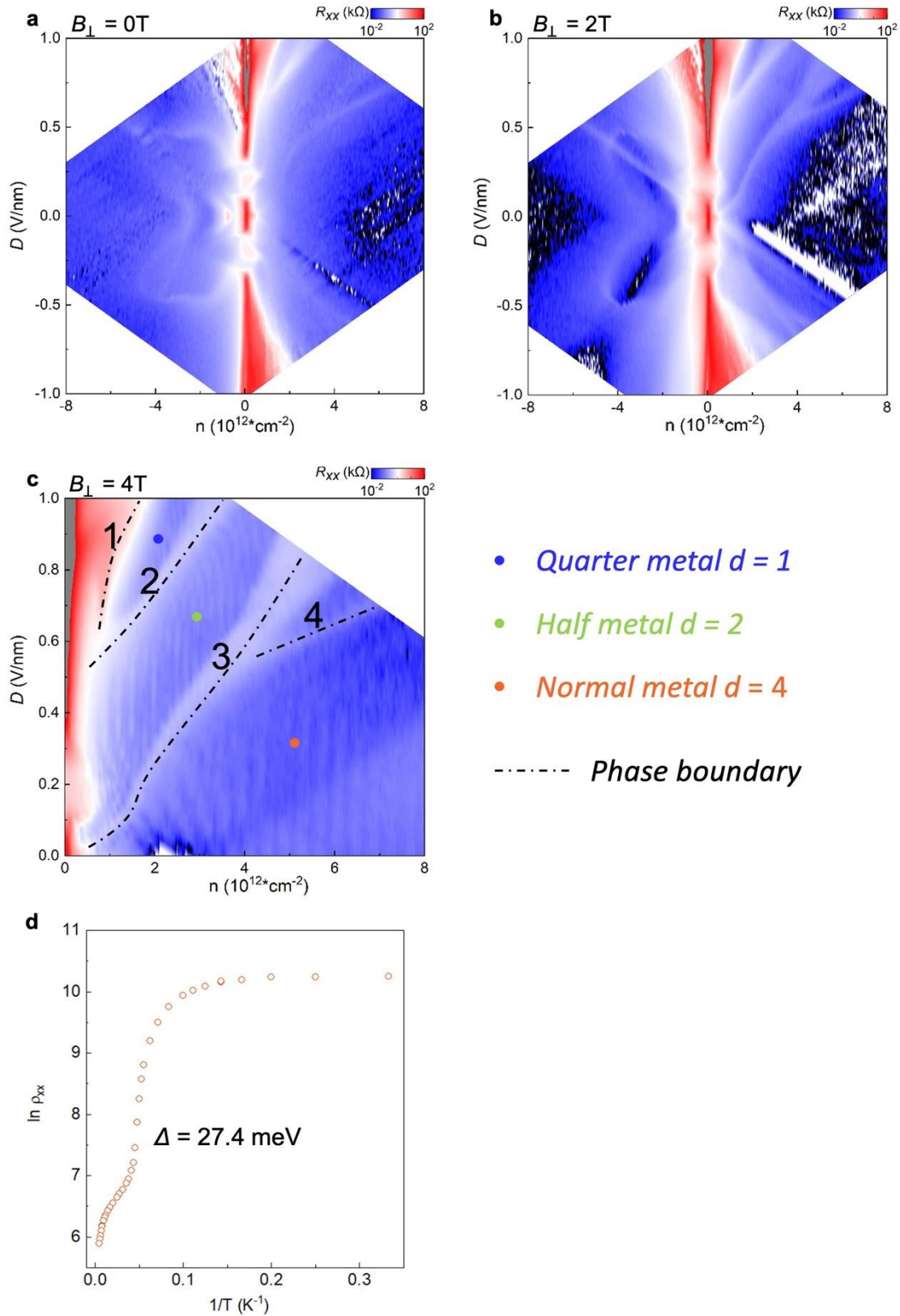

**Extended Data Fig. 3. Transport measurements of *r*-6LG without moiré under different magnetic field $B_\perp$. a, b, c,** Color plots of the four-probe resistance $R_{xx}$ as functions of the filling factor $v$ and $D$ at $B_\perp$ = 0 T, 2T, and 4T, respectively. Black

dashed lines indicate the phase boundaries. Colored dots label the different phases, including spin and valley polarized quarter metal, spin-polarized half metal and normal metal. The parameter $d$ denotes the degeneracy. d, Arrhenius plot of ln ($R_{xx}$) versus $T^{-1}$ at $v = 0$ and $D = 0$ V/nm. The insulating gap $\Delta = 27.4$ meV is extracted from the thermal activation equation $R_{xx} \propto e^{-\Delta/2k_BT}$.

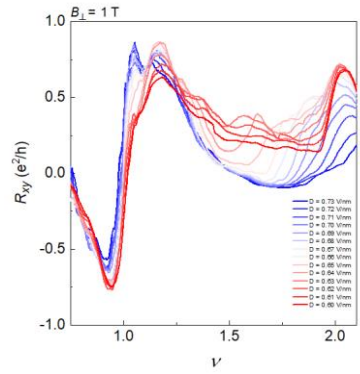

**Extended Data Fig. 4.** Line plots of transverse resistance $R_{xy}$ at different $D$ fixed at $B_\perp = 1$ T

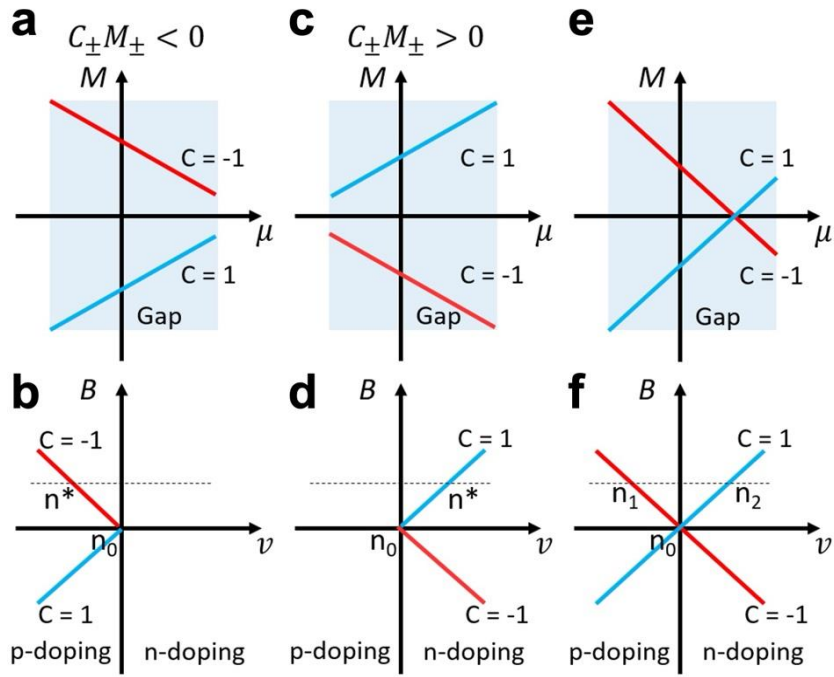

**Extended Data Fig. 6. Relationship between the sign of the Chern insulator and magnetization $M(\mu)$**[31]**. a, c, e,** The sign of $M(\mu)$ across the gap. **b, d, f,** The Chern insulators with different sign. If $M(\mu)$ does not change sign across the gap, the Chern insulator cannot change sign, resulting in a Chern insulator with only have one Chern number, either positive or negative. If $M(\mu)$ changes sign across the gap, the sign of the Chern insulator can be reversed by doping.